\documentclass[]{natureprintstyle}
\usepackage{graphicx,amssymb,amsmath}
\usepackage{epstopdf}
\usepackage{verbatim}
\usepackage{multirow}
\usepackage{hyperref}
\usepackage[T1]{fontenc} 
\usepackage[utf8]{inputenc}
\hypersetup{
    colorlinks=true,       		
    linkcolor=black,          	
    citecolor=black,            
    filecolor=black,      		
    urlcolor=black,           	
    runcolor=cyan
}

\DeclareSymbolFont{AMSb}{U}{msb}{m}{n}
\DeclareSymbolFontAlphabet{\mathbb}{AMSb}

\begin{document}

\title{\LARGE A  solenoidal synthetic field and the non-Abelian Aharonov-Bohm effects in neutral atoms}

\author{\large Ming-Xia Huo$^{1}$, Wei Nie$^1$, David A.W. Hutchinson $^{1,2}$, and Leong Chuan Kwek $^{1,3,4}$}

\maketitle

\begin{affiliations}
\item Centre for Quantum Technologies, National University of Singapore, 3 Science Drive 2, Singapore 117543\\
\item Department of Physics, University of Otago, Dunedin, New Zealand\\
\item National Institute of Education, Nanyang Technological University, 1 Nanyang Walk, Singapore 637616\\
\item Institute of Advanced Studies, Nanyang Technological University, 60 Nanyang View, Singapore 639673\\ 
\end{affiliations} 


\textbf{
Cold neutral atoms provide a versatile and controllable platform for emulating various quantum systems.
Despite efforts to develop artificial gauge fields in these systems, realizing a unique ideal-solenoid-shaped magnetic field within the quantum domain in any real-world physical system remains elusive.
Here we propose a scheme to generate a "hairline" solenoid with an extremely small size around 1 micrometer which is smaller than the typical coherence length in cold atoms.
Correspondingly, interference effects will play a role in transport.
Despite the small size, the magnetic flux imposed on the atoms is very large thanks to the very strong field generated inside the solenoid.
By arranging different sets of  Laguerre-Gauss (LG) lasers, the generation of Abelian and non-Abelian SU($2$) lattice gauge fields is proposed for neutral atoms in ring- and square-shaped optical lattices.
As an application, interference patterns of the magnetic type-I Aharonov-Bohm (AB) effect are obtained by evolving atoms along a circle over several tens of lattice cells.
During the evolution, the quantum coherence is maintained and the atoms are exposed to a large magnetic flux.
The scheme requires only standard optical access, and is robust to weak particle interactions.
}

There is intense interest in finding new settings in which different forms of gauge fields appear.
Ultracold atoms in optical lattices are ideal systems with which to achieve this goal as they offer unprecedented possibilities of emulating condensed matter and high energy physics models~\cite{Dalibard,Lewenstein1,Lewenstein2,Juzeliunas1,Juzeliunas2,Juzeliunas3,Lin1,Lin2,Osterloh1,Osterloh2,Osterloh3,Osterloh4,Klein,Banerjee1,Banerjee2,Banerjee3,Gorshkov1,Gorshkov2,
Gorshkov3,Gorshkov4,Gorshkov5,Gorshkov6}.
The original magnetic and electric AB effects (type-I) are distinguished from the neutral scalar AB effect and the Aharonov-Casher effect (type-II) by the absence of any electromagnetic fields~\cite{AB}.
This original AB effect shows that electron wave packets are influenced by the non-zero potentials and can obtain a non-zero phase shift although they travel in field-free regions~\cite{AB}.
The type-I AB effect is important conceptually because it bears on three issues: whether potentials are physical quantities, whether action principles are fundamental, and the principle of locality.
As such it has spawned many studies~\cite{Cham,Toromura,Caprez,Shinohara,Webb1,Webb2,Webb3,Webb4,Webb5,Russo1,Russo2,Bayer,Ribeiro1,Ribeiro2,Adrian1,Adrian2}. 
In 1986, to overcome the issue of stray fields in the previous experiments, Toromura \text{et al.} have performed a beautiful experiment to prove the presence of the AB phase by using magnetic toroids with superconducting shields to eliminate the leakage fields~\cite{Toromura}.
In 2007, Adam \textit{et al.} tested the magnetic type-I AB effect in the absence of any force for a macroscopic system~\cite{Adam}.
Although the phase shift has been observed experimentally, the direct interference signatures in the original version of the AB effect have not yet been reported.
In solid-state AB interferometer, the AB effect has also been nicely demonstrated~\cite{Yang}.
Given the great importance of type-I AB effect as a quantum mechanical phenomenon, it is of fundamental interest to ask whether the original AB effect can be reproduced in a quantum regime and how to observe the associated interference effects.
A good realization in degenerate cold gases has however so far proved elusive. 

\begin{figure}
\centering
\includegraphics[scale=.7]{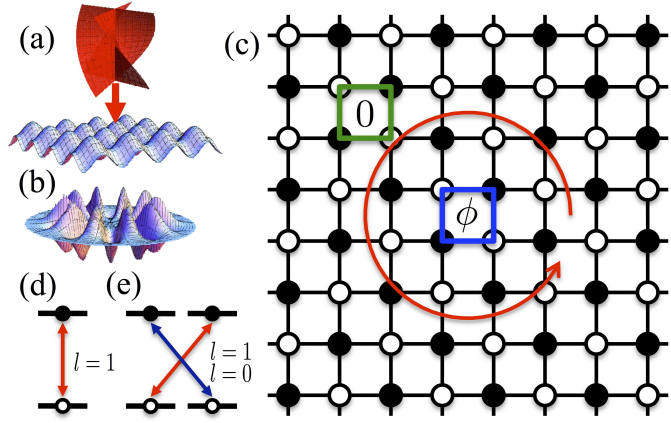}
\caption{
Schematic diagram for generating strongly localized effective Abelian and non-Abelian gauge fields with cold atoms trapped in square (a) and ring (b) lattices.
The atoms can hop from one site to a neighbouring site via laser-assisted tunnelling by shinning LG lasers perpendicular to the lattice surface (a).
The LG beams drive the spin-conserved transitions between two identical sublevels in neighbouring sites in (d) to generate Abelian fields, and spin-flipping transitions between two different sublevels in neighbouring sites in (e) to generate SU($2$) non-Abelian fields.
The resulting magnetic flux as shown in (c) is non-zero when going around a loop (e.g., the blue square) including the laser centre which is addressed in the lattice centre and zero when the laser centre is excluded from the loop (e.g., the green square).
}
\label{fig1}
\end{figure}

For the simulation of uniform or staggered gauge fields, several studies have proposed a number of methods~\cite{Banerjee1,Banerjee2,Banerjee3,Gorshkov1,Gorshkov2,Gorshkov3,Gorshkov4,Gorshkov5,Gorshkov6}, including rotation of the confining trap~\cite{Dalibard,Lewenstein1,Lewenstein2}, using adiabatic motion of multilevel atoms in a dark-state picture~\cite{Juzeliunas1,Juzeliunas2,Juzeliunas3}, and employing spatially dependent two-photon dressing~\cite{Lin1,Lin2}. 
In the presence of a lattice potential, schemes to generate gauge fields have been proposed using staggered-aligned states~\cite{Osterloh1,Osterloh2,Osterloh3,Osterloh4} or through rotations~\cite{Klein}.
In this work, we propose a scheme to emulate ideal-solenoid-shaped Abelian and non-Abelian gauge fields by employing  LG lasers~\cite{Aidelsburger}.
The generated magnetic field is strongly localized within a lattice cell, which allows an initially prepared coherent state of atoms to evolve along a circle without any electromagnetic field.
Due to the small radius of the solenoid, the atoms can evolve along two paths within coherence time and eventually overlap so that interference occurs. At the same time, the strong strength of the magnetic field exposes the atoms to a large magnetic flux which results in a clearly visible interference signature.
Besides mirroring the original AB effect relating to an Abelian gauge field, we also investigate systems exposed to a non-Abelian field which is achievable by employing more lasers.
Different from the AB phase mentioned in other proposals which is accumulated from atomic evolution along four sides of a lattice cell, the AB phase here is accumulated from atomic evolution along a circle enclosing several tens of lattice cells, which makes the detection of interference more readily accessible to experiments. 

\bigskip

\section*{Results \\} 

In this work, we consider a scheme to realize a "hairline" solenoid in both a square lattice [Fig.~\ref{fig1}(a)] and in a ring geometry~\cite{Amico} [Fig.~\ref{fig1}(b)] by employing LG-laser-assisted tunnelling.
Our scheme consists of three ingredients.
Firstly, atoms in different internal states are trapped in a staggered manner in a ring or a square lattice, e.g., as shown in Fig.~\ref{fig1}(c) for a square-lattice case.
The open and solid circles represent atoms in different internal states, respectively.
Secondly, the depth of lattice potentials is tuned to be very large such that regular tunnelling among lattice sites is prohibited.
Finally, the additional lasers with non-zero orbital angular momentum are switched on to induce tunnelling between the adjacent lattice sites.

In particular, we consider a trapping scheme with the basic principle proposed in previous works~\cite{Osterloh2,Osterloh3}, where two-electron atoms generally possess a spin-singlet ground state $g$ and a long-lived spin triplet excited state $e$ with a lifetime around $20$ $\mathrm{s}$  \cite{Osterloh3,Porsev}. The $g$ and $e$ states have opposite polarizabilities at the \textquotedblleft anti-magic\textquotedblright\ wavelength $\lambda\simeq 1 \mathrm{\mu m}$ \cite{Osterloh3}, where they experience potentials with opposite signs and therefore align in a staggered manner, as shown in Fig.~\ref{fig1}(c) for the square-lattice case. We would like to note that apart from the typical square lattice system which can be created by the interference of counter-propagating laser beams at the \textquotedblleft anti-magic\textquotedblright\ wavelength, a ring lattice in (b) can also be created by interfering an off-resonant LG laser and a plane wave with their wavelengths chosen to be at the \textquotedblleft anti-magic\textquotedblright\ wavelength~\cite{Amico}. At this stage, the lattice depth is set to be very large such that the tunnelling of atoms is strongly suppressed. As shown in Fig.~\ref{fig1}(d), resonant LG lasers are applied to drive transitions between the $g$ and $e$ states, leading to atomic hoppings along a loop as shown in Fig.~\ref{fig1}(c). Different values of $l$ correspond to different accumulated phase when atoms move around the loop. We consider $l=1$ and $l=0$ lasers in this work. For the former with $l=1$, it is found that the accumulated phase is $\pi$, which resembles a system where an Abelian magnetic field is applied to the atoms. For the latter with $l=0$, the accumulated phase is $0$, thus emulating a system with a zero magnetic field applied to the atoms. Moreover, when considering two sub-states in each $g$ and $e$ levels, which is shown in Fig.~\ref{fig1}(e), an SU($2$) field can be generated by employing two LG lasers with $l=0$ and $l=1$ to drive transitions between different sub-states.

We first consider a Hamiltonian describing atoms with only one sub-state in each $g$ and $e$ levels as
\begin{eqnarray}
H&=&\sum_{s=g,e}\int d\mathbf{r} \psi^{\dagger}_s(\mathbf{r}) [\frac{\mathbf{\hat{p}}^2}{2m}+\eta_s V(\mathbf{r})] \psi_s(\mathbf{r})
-\int d\mathbf{r} [d_{eg}E(\mathbf{r})^* e^{i\omega t} \psi^{\dagger}_e(\mathbf{r})\psi_g(\mathbf{r})+h.c.] \notag \\
&&+\sum_{s,s'=g,e}g_{s,s'}\int d\mathbf{r} \psi_s^{\dagger}(\mathbf{r})\psi_{s'}^{\dagger}(\mathbf{r})\psi_{s'}(\mathbf{r})\psi_s(\mathbf{r}).
\end{eqnarray}
The four terms in turn give the kinetic energy, the lattice potential, the laser-assisted transition, and the interaction between atoms, respectively.
Here, $\psi_{s}(\mathbf{r})$ represents the atomic field operator at position $\mathbf{r}$, $\mathbf{\hat{p}}$ is the momentum operator, and $m$ is the mass.  The state-dependent sign of the lattice potential $V(r)$ is denoted by $\eta_g=+$ and $\eta_e=-$, due to the lattice lasers at the "anti-magic" wavelength. The dipole moment of the $g$-$e$ transition is labelled as $d_{eg}$, and the interaction strength is given by $g_{s,s'}$. 
As shown in Fig. \ref{fig1}(a), the LG laser is applied perpendicular to the lattice. The amplitude of the LG laser resonant to  the $g$-$e$ transition reads
\begin{eqnarray}
E(\mathbf{r}) = E f_{pl}(r) e^{il\varphi} e^{i(\omega t-kz)},
\end{eqnarray}
where $f_{pl}(r) = (-1)^{p} \sqrt{\frac{2p!}{\pi (p+|l|)!}} \xi^{|l|+2} L_{p}^{|l|} e^{-\xi^{2}}$,
$\xi =\frac{\sqrt{2}r}{r_{\mathrm{w}}}$, $r_{\mathrm{w}}$ is the waist of the beam, and $L_{p}^{|l|}$ are the Laguerre functions~\cite{Amico}. For an $N_{S} \times N_{S}$ 2D lattice, we choose $r_{\mathrm{w}}=N_{S}a/2$. 
The cylindrical coordinate ($r$, $\varphi$, $z$) is chosen that the longitudinal axis $z$ is along the propagation direction of the LG laser, and the labels $p$ and $l$ represent the radial and azimuthal quantum numbers, respectively.

We choose a lattice potential $V$ that is minimized at sites $\mathcal{G}=\{\text{open-circle sites}\}$ and maximized at sites $\mathcal{E}=\{\text{solid-circle sites}\}$ (see Fig.~\ref{fig1}).
In the presence of a deep lattice potential, $\psi_s(\mathbf{r})$ can be expressed by Wannier functions in the lowest band as $\psi_{g(e)}^{\dagger}(\mathbf{r})\simeq \sum_{j\in\mathcal{G(E)}}a_{j}^{\dagger}\omega^{*}(\mathbf{r}-\mathbf{r_{j}})$, where $\omega(\mathbf{r}-\mathbf{r}_{j})$ is the Wannier function \cite{Jaksch}. In the numerical simulation, we approximate the Wannier functions to be Gaussion functions.
Here, we have assumed that the lattice potential is symmetric for the $g$ and $e$ states.
The Hamiltonian then reduces to a tight-binding model as
\begin{eqnarray}
H=-\sum_{\langle i,j \rangle}(J_{i,j}a_{i}^{\dagger}a_{j}+h.c.)
-J_{0}\sum_{\langle\langle i,j \rangle\rangle}(a_{i}^{\dagger}a_{j}+h.c.)
+\sum_{i}\epsilon_{i}a_{i}^{\dagger}a_{i}
+\frac{U}{2}\sum_{i}a_{i}^{\dagger}a_{i}^{\dagger}a_{i}a_{i}.
\end{eqnarray}
Here, the chemical potential $\epsilon_{i}$ is not relevant in terms of the gauge fields, so we treat it as uniform by tuning the lattice potential. We assume the on-site interaction strength to be identical for each site and label it as $U$.
For the first hopping term, $\langle i,j \rangle$ denotes two nearest-neighbouring sites that one is in the set $\mathcal{G}$ while the other is in the set $\mathcal{E}$, between which the tunnelling is induced by the LG laser and has a strength~\cite{Osterloh2,Osterloh3}
\begin{eqnarray}
J_{i,j}=d_{eg}E \int d\mathbf{r} \omega^{*}(\mathbf{r}-\mathbf{r}_i) f_{pl}(r) e^{-il\varphi} e^{ikz} \omega(\mathbf{r}-\mathbf{r}_{j}),
\label{J}
\end{eqnarray}
where $i\in \mathcal{G}$ and $j\in \mathcal{E}$.
For the second hopping term, $\langle\langle i,j \rangle\rangle$ denotes two next-nearest-neighbouring sites, which are in the same set $\mathcal{G}$ or $\mathcal{E}$, and the corresponding tunnelling strength $J_0=-\int d\mathbf{r} \omega^{*}(\mathbf{r}-\mathbf{r}_i) [\mathbf{p}^2/2m+\eta_s V(\mathbf{r})]\omega(\mathbf{r}-\mathbf{r}_{j})$. 
By controlling the lattice potential, the next-nearest-neighbouring tunnelling $J_0$ can be tuned to be much smaller than the nearest-neighbouring $J_{i,j}$, which allows us to neglect $J_0$ and only focus on $J_{i,j}$ in the following.

A natural property of the laser with a non-zero orbital angular momentum is a spatial-dependent phase.
Since the laser is applied perpendicular to the lattice, the phase of $J_{i,j}$ will only depend on the azimuthal angle $\varphi$, as shown in Eq. (\ref{J}).
The phase of $J_{i,j}$ can be estimated as $J_{i,j}\propto e^{-il\varphi}$ with $\varphi$ the azimuthal angle of the midpoint between two adjacent sites $i\in \mathcal{G}$ and $j\in \mathcal{E}$.
We would like to remark that the sign of the phase depends on the tunnelling direction, where for $i\in \mathcal{E}$ and $j\in \mathcal{G}$, $J_{i,j} \propto e^{il\varphi}$.
Due to the vortex of the LG laser, the accumulated phase for atoms moving around a loop enclosing or excluding the laser centre are different.
As a simplified illustration, we consider a centre cell of the square lattice [blue square in the centre of Fig. \ref{fig1} (c)].
Here, we assume that the laser centre coincides with the centre of this cell.
As a four-site circle, $\mathcal{G}$ and $\mathcal{E}$ sites appear alternatively.
Without the loss of generality, we assume $\varphi = 0$ along the upward direction in Fig.~\ref{fig1} (c).
Then, for a particle moving in a counterclockwise direction, it undergoes tunnelling with phases $0$, $\frac{l\pi}{2}$, $-l\pi$, and $\frac{3l\pi}{2}$ in the four links, respectively.
As a result, the accumulated phase around the centre cell is given by $0+\frac{l\pi}{2}-l\pi+\frac{3l\pi}{2}=l\pi$.
By choosing an odd $l$, the accumulated phase is nontrivial, giving a non-zero gauge field within the centre cell.
In the following, we will focus on the case with $l=1$ for the generation of an Abelian gauge field. As a comparison, a laser with $l=0$ will give a zero accumulated phase and therefore correspond to a zero gauge field. 
On the other hand, for a cell far away from the laser centre [green square in Fig.~\ref{fig1} (c)], the accumulated phase is always zero for any $l$, which means that a non-zero gauge field is strongly localized inside the centre cell for $l=1$, resembling a very thin solenoid. As an illustration, we consider a cell centred around $\mathbf{r}_{0}$. The laser centre is set to be the origin of the system. The coordinates of each site belonging to the cell is $\mathbf{r}=\mathbf{r}_{0}+\delta \mathbf{r}$, which gives the arimuthal angle
\begin{eqnarray}
\varphi=\mathrm{arctan}(\frac{y}{x})\simeq \varphi_{0}+\frac{\mathrm{cos}\varphi_{0}\delta y-\mathrm{sin}\varphi_{0}\delta x}{r_{0}}, 
\end{eqnarray}
provided $r_{0}\gg a$. Here $\varphi_{0}$ is the azimuthal angle of $r_{0}$, and $a$ is the lattice constant. With $\delta x, \delta y= \pm \frac{a}{2}$, the accumulated phase is zero for a cell away from the laser centre.

\begin{figure}
\centering
\includegraphics[scale=.8]{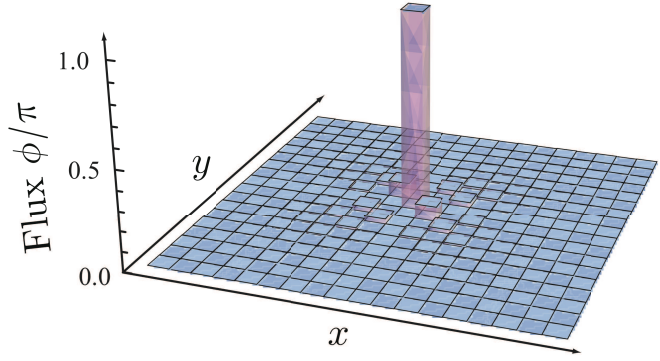}
\caption{
Numerically calculated net flux for each cell on a 2D square lattice.
We have chosen $l=1$ for the LG laser, and the laser centre is addressed in the lattice centre.
The flux is $\pi$ when atoms tunnel along a loop including the laser centre.
Although small fluctuations appear around the centre cell, the positive and negative phases cancel each other and the total flux for a loop including the centre cell remains a value of $\pi$.
}
\label{fig2}
\end{figure}

Numerical simulation for the accumulated flux $\phi_{ij}$ is shown in Fig.~\ref{fig2}, where an $l=1$ LG laser drives the tunnelling of atoms along a loop enclosing the lattice centre that coincides with the laser centre. The flux is defined as $\phi_{i, j}=\text{arg}(J_{i, j})-\text{arg}(J_{i, j+1})+\text{arg}(J_{i+1, j+1})-\text{arg}(J_{i+1, j})$, where the tunneling strength $J_{i, j}$ is given in Eq. (\ref{J}). As shown in Fig.~\ref{fig2}, the gauge field is non-zero within only a few cells around the centre. We would like to remark that even if the centre of the laser is slightly shifted from that of the lattice, it will not affect the results significantly, although some small fluctuations may appear around the solenoid. 
In a similar way, an LG laser drives tunnelling of atoms which are trapped in a ring lattice. For an $l=1$ laser, the accumulated phase along the ring is $\pi$. It is straightforward to extend to the SU(2) case, where $\psi_{s}$ becomes a $2 \times 1$ column matrix to include the two sub-states $g_{1}$ and $g_{2}$ ($e_{1}$ and $e_{2}$) in $g$ ($e$) level [see Fig. 1(e)]. Two laser beams with $l=0$ and $l=1$ are employed to drive $g_{1}$-$e_{2}$ and $g_{2}$-$e_{1}$ transitions, respectively. To give a unitary hopping matrix for two spins, we should choose $l=0$ and $l=1$ lasers with the same amplitude to drive the spin-flipping transitions. It is satisfied when we apply an LG mode with $p=0$, $l=1$, and a superposition of two LG modes with $p=0$, $l=0$\ and $p=1$, $l=0$. The non-Abelian tunneling matrix is then given by
\begin{equation}
\hat{J}_{ij}= \left( \begin{array}{ccc}
0 & \vert J_{ij}\vert \\
J_{ij} & 0 \end{array} \right), 
\end{equation}
where $J_{ij}$ is given in Eq. (\ref{J}).

\begin{figure}
\centering
\includegraphics[scale=.7]{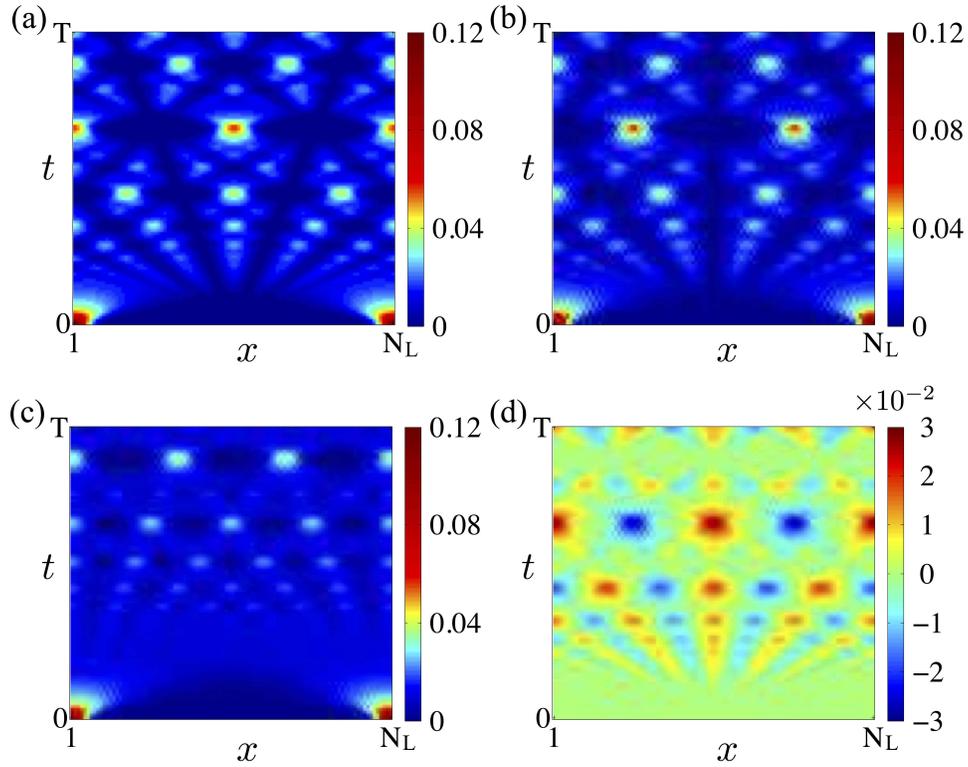}
\caption{The time evolution of particle distributions for particles hopping
around a loop formed by LG beams in a ring with $N_{\mathrm{L}}=100$ sites
under a zero gauge field (a), an Abelian U(1) field (b), or a non-Abelian
SU($2$) field ((c) and (d)). The particle densities are represented by the
percentages of particle numbers to the total number of the initially
prepared Bose-Einstein condensate. At time $t=0$, the particles are prepared
around the site $x=1$. During the time period from $t=0$ to $t=T=600\hbar/E_{R}$,
the fringe patterns around the opposite site $x=N_{\mathrm{L}}/2$ become
completely different for the three field cases ($B=0$, U(1) and SU($2$)). 
Around this site, there is always a constructive interference for the $l=0$ case in (a) and a
destructive interference for $l=1$ in (b). In contrast, for the non-Abelian case, the
charge wave density,  the sum of the two effective spins, has no unique behavior
as shown in (c), and the spin wave density (d), as a difference of the two effective
spins, exhibits the red-color (blue-color) component with a constructive (destructive) interference around
the site $x=N_{\mathrm{L}}/2$ and a destructive (constructive) interference
around the nearest neighboring area. Movies of the time evolution are available in the supplementary materials. }
\label{fig3}
\end{figure}

We would like to note the difference between our scheme and the works for continuous systems investigated in Refs. [13,14].
Although both employ laser beams with orbital angular momentum, the latter schemes deal with dark states in the electromagnetically induced transparency configuration, where the adiabatic motion of multilevel atoms is applied. In the present scheme, the necessary condition for the generation of gauge fields is a staggered distribution of different internal states and laser-assisted tunnelling. Therefore, unlike a uniform field generated in Refs. [13,14],
a solenoidal field is produced here. 

\begin{figure}
\centering
\includegraphics[scale=.7]{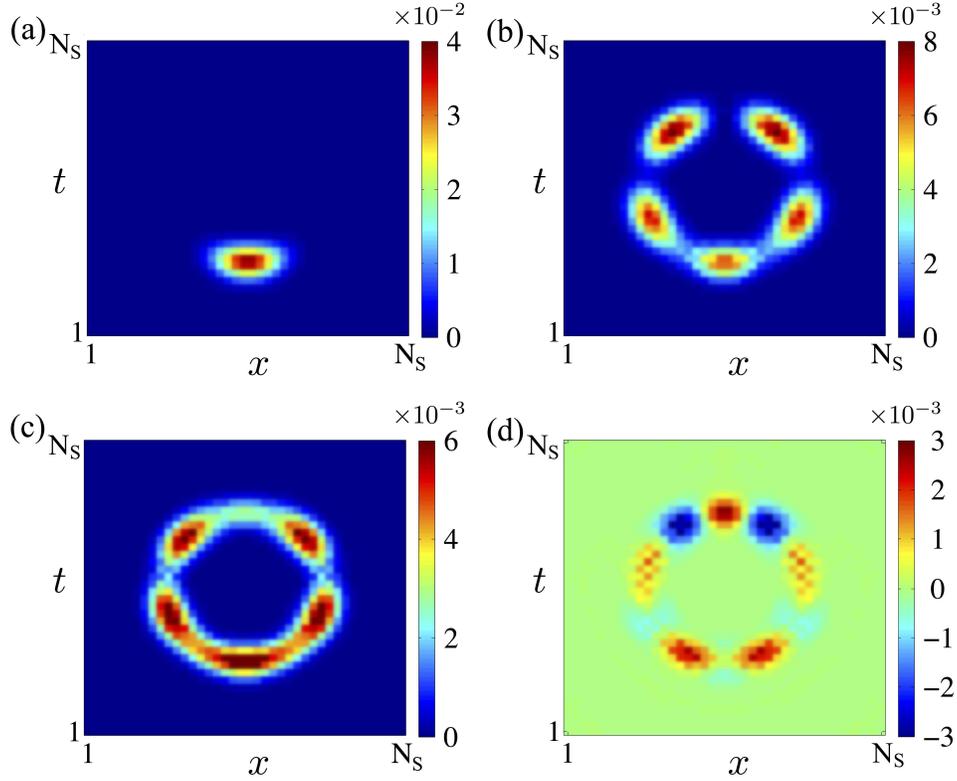}
\caption{The time evolution of particle distributions for particles hopping
around a loop formed by LG beams in a 2D square lattice with $40\times 40$
sites. At time $t=0$ as shown in (a), the particles are distributed around 
one site. The screenshots at time $t=4000\hbar/E_{R}$ are given for the system under an Abelian
U(1) (b) or a non-Abelian SU($2$) ((c) and (d)) gauge field. The
particle densities are represented by the percentages of particle numbers to
the total number of the initially prepared Bose-Einstein condensate. While a
destructive interference always occurs around the opposite site in the
Abelian field case (b), a non-vanishing charge wave density occurs
everywhere in (c) and spin waves appear in (d) with red- and
blue-colored components, where the red-color (blue-color) component has a
constructive (destructive) interference at the opposite side and a
destructive (constructive) interference at its two neighboring area. Movies of the time evolution are available in the supplementary materials.}
\label{fig4}
\end{figure}

\noindent \textit{Type-I AB  effect}: To see the interference patterns, we load a Bose-Einstein condensate (BEC) initially away from the centre of the LG laser, as seen in the plot shown in Fig. \ref{fig3}(a)-(c) at $t=0$ for a ring-lattice case and in Fig. \ref{fig4}(a) for a square-lattice case. The system is then allowed to evolve under zero, Abelian and non-Abelian gauge fields. And finally the fringe patterns are measured with time. After the initial BEC state is prepared, the LG beams are switched on. At this stage, the Hamiltonian governing the time evolution is 
\begin{eqnarray}
H=-\sum_{\langle i,j \rangle}(J_{i,j}a_{i}^{\dagger}a_{j}+h.c.).
\end{eqnarray}
Here we have neglected the interactions between atoms. Evolutions for  weakly interacting gas are analyzed in the Methods section, where the results show that the key features of the 
interference patterns remain robust against a weak interaction. 
For an LG laser with $l=1$, the accumulated phase along the circle is $\pi$. Since the atoms move along left or right path, the atoms from different paths will possess a different phase factor at the opposite site, giving a destructive interference. For an LG laser with $l=0$, the phase is always zero, which resembles the system with particles moving in the absence of any gauge field. The atoms evolving along two different paths will possess the same phase factor at the opposite site, and the interference is destructive. Therefore, the interference fringes are closely related to the phase in our scheme. 

For a 1D ring-shaped optical lattice as
illustrated in Fig. \ref{fig1}(b), numerical simulations for spatial
distributions of particle numbers are plotted in Fig. \ref{fig3}, where the
particle numbers are shown as a percentage of the total number in the
initially prepared BEC. We have chosen the lattice site
number as $N_{\mathrm{L}}=100$\ and the time period $T$\ as $600$ in 
unit of $E_{R}/\hbar$, where $E_{R}$ is the recoil energy and the typical hopping is $J\simeq 0.05E_{R}$. 
With $E_{R}/\hbar = 2\pi \times 900$ $\mathrm{Hz}$ \cite{Osterloh3}, the time scale is approximately in the range of $100$ $\mathrm{ms}$. 
Distinctly different interference pattens can be seen around
$x=N_{\mathrm{L}}/2$ site for particles experiencing a zero magnetic field as shown in (a), an Abelian
U(1) gauge field as shown in (b), and a non-Abelian SU($2$) gauge field as
shown in (c)-(d). The reason for the two subfigures in the SU($2$) case is
that two effective spins are involved in the SU($2$) field. We plot
an effective charge wave density (the sum of densities of the two effective
spins) in (c) and an effective spin wave density (the difference of
densities of the two effective spins) in (d), respectively. 
At the $N_{\mathrm{L}}/2$ site, a destructive (constructive) interference is always seen in (b) [(a)]. 
Two components appear with one colored red and the other blue, where the constructive interference occurs
for one type (red). 

For the 2D square lattice case as illustrated in Fig. \ref{fig1}(a) and \ref%
{fig1}(c), we can also prepare the initial state by loading a BEC around one
plaquette. 
For the U(1) Abelian field case as shown in Fig. \ref{fig4}(b), we turn on
an $l=1$ laser as illustrated in Fig. \ref{fig1}(d), and for the non-Abelian
SU($2$) field case as shown in Fig. \ref{fig4} (c)-(d), we turn on two
orthogonally polarized $l=0$ and $l=1$\ lasers as illustrated in Fig. \ref%
{fig1}(e). Numerical simulations with time show that the angular momentum
imparted by the LG beams results in a circular distributions of atoms. The
particle number distribution at the opposite side of the circle from the
initially prepared site again shows clear destructive interference in the
case of the Abelian gauge field - the signature of the AB effect. As a
concrete example, screenshots at time $t=4000\hbar/E_{R}$ are given in Fig. %
\ref{fig4} (b)-(d). The effect is similar to the ring geometry case. 

Our key result for the ring geometry is a clear demonstration of the magnetic type-I AB
effect, where in the presence of an Abelian gauge field, the AB phase
always induces destructive interference at the $N_{\mathrm{L}}/2$ site. Absence
of any population at the $N_{\mathrm{L}}/2$ site at any time is therefore a
clear signature of the Abelian AB effect. Furthermore, we demonstrate the
effects of the non-Abelian gauge field in terms of the spin density waves.

\section*{Discussion \\}
We have presented a method to generate a "hairline" solenoid with an extremely small size around 1 micrometer. 
Clear evidences of AB effect in the interference patterns for both Abelian and non-Abelian SU($2$) fields are obtained with a unique topological 
structure in both a ring geometry and a square lattice.  
A maximal magnetic flux is achieved as well as quantum coherence maintained during the evolution process of atoms due to 
the strong strength of generated magnetic field and the small radius of the solenoid. In the detection area, the interference 
is always constructive for the case of zero field and always destructive for the Abelian field case. For 
the non-Abelian field case, the spin wave density exhibits two components with a constructive or 
destructive interference in the detection area. The scheme requires standard optical access within the reach of current techniques, and the key conclusions are not 
affected by the weak particle interactions. Moreover, these effects are robust against decoherence as they are topological in origin. 
Consequently, they could in principle be harnessed for quantum information processing \cite{Leek,Jones,Falci}. The technique is 
also straightforwardly generalized to SU($N$) gauge fields.



\section*{Methods \\} 

\noindent \textit{Interaction effect}: In the main text, we have considered the non-interacting gas. A weakly interacting degenerate Bose gas can be described in the framework of the Gross-Pitaevskii equation, and now we will treat our models at the Gross-Pitaevskii level. To start with, we write down the Hamiltonian that describes interacting bosons evolving on a lattice and subjected to a gauge field:
\begin{eqnarray}
H=-\sum_{\langle i,j \rangle}(J_{i,j}a_{i}^{\dagger}a_{j}+h.c.)
+\frac{U}{2}\sum_{i}a_{i}^{\dagger}a_{i}^{\dagger}a_{i}a_{i}.
\end{eqnarray}
Here $a_{i}$ is a canonical Bose annihilation operator on sites of the optical lattice labeled by the integer $i$, $J_{ij}$ is the hopping amplitude, and $U$ is the interaction strength. For a weakly interacting gas, which is the typical case for the atomic gas \cite{Zwerger}, the low temperature dynamics of $H$ can be described by introducing the complex dynamical mean field variable $\psi_{i}(t)$ whose value is a measure of $\langle a_{i}(t) \rangle /\sqrt{\langle n\rangle}$. Here $\langle n \rangle$ is the average atom occupancy per lattice site. The dynamics is then described by the discrete Gross-Pitaevskii equation \cite{Polkovnikov}
\begin{equation}
i\hbar\frac{\partial }{\partial t}\psi_{i}=\sum_{i} [-(J_{i,i+1}\psi_{i+1}+J_{i,i-1}\psi_{i-1})
+\lambda J\vert {\psi_{i}} \vert ^{2 } \psi_{i}],
\end{equation}
where 
$\lambda=\frac{U\langle n\rangle}{2J}$. 
By numerically solving the differential equations for evolutions driven by systems exposed to zero, Abelian and non-Abelian gauge fields, we have shown that the key features of the interference patterns presented here remain robust up to at least $\lambda=5$, which is easily achievable within current experimental setups. 

\begin{addendum}

\item [Acknowledgements] 
The authors are grateful for the financial support of the National Research Foundation \& Ministry of Education, Singapore.

\item [Author contributions] 
All authors M.-X.H., W.N., D.A.W.H. and L.-C.K. contributed equally to the results.

\item [Additional information] 
Supplementary information is available in the \href{www.nature.com/}{online version of the paper}. Reprints and
permissions information is available online at \href{www.nature.com/reprints}{www.nature.com/reprints}.
Correspondence and requests for materials should be addressed to
huo.mingxia@physics.ox.ac.uk, david.hutchinson@otago.ac.nz, kwekleongchuan@nus.edu.sg.

\item [Competing financial Interests] 
The authors declare no competing financial interests.

\end{addendum}

\clearpage

\end{document}